\documentstyle[epsfig,color]{elsart}
\begin{document}

\begin{frontmatter}
\title{On the chemical equilibration of strangeness-exchange reactions
in heavy-ion collisions}

\author[UCT]{J.~Cleymans},
\author[TUD,CERN]{A.~F\"orster},
\author[TUD,CERN]{H.~Oeschler},
\author[UW]{K.~Redlich},
\author[TUD]{F.~Uhlig}
\address[UCT]{Centre for High Performance Computing and Department  of  Physics,  University of Cape Town,\\
Rondebosch 7701, South Africa}
\address[TUD]{Institut f\"ur Kernphysik, 
Technische Universit\"at Darmstadt, D-64289~Darmstadt, Germany}
\address[UW]{Institute of Theoretical Physics, University of Wroc\l aw,
 Pl-45204 Wroc\l aw, Poland }
\address[CERN]{CERN, CH-1211 Geneva 23, Switzerland}

\begin{abstract}
The strangeness-exchange reaction $\pi + Y \leftrightarrow K^- + N $
is shown to be the dynamical origin of chemical saturation and equilibration
for $K^-$ production 
in heavy-ion collisions up to  beam energies of 10 $A\cdot$GeV.
The hyperons occurring in this process are produced associately
with $K^+$ in baryon-baryon and meson-baryon interactions. This
connection is demonstrated by the ratio of $K^-/K^+$ which does
not vary with centrality and shows a linear correlation with the
yield of pions per participant.
 At incident energies above AGS 
this  correlation    no longer holds due to the change in the
production mechanism of kaons.

\end{abstract}
\end{frontmatter}

A fairly complete set of experimental results on $K^-$ production
in heavy-ion collisions for beam energies from 1.5 $A\cdot$GeV up
to about 10 $A\cdot$GeV  has now become available from
SIS~\cite{Barth,Laue,Menzel,wisniewski,AF} and from
AGS~\cite{Ahle_1998,Ahle_2000,Klay}. These results have
attracted a lot of interest as the yield of $K^-$
compared to the yield of $K^+$ is significantly higher in
heavy-ion collisions than in elementary $N$--$N$ collisions
\cite{Barth,Laue,Menzel}. Transport model
calculations~\cite{cassing} lead to the
interpretation that this effect may be caused by an attractive
$K^-N$ potential.
Alternatively, thermal-statistical models  using bare particle
masses were also shown to be very successful in describing particle
yields (including the $K^-$ one) down to low incident energies 
around 1.5 $A\cdot$GeV~\cite{CLE99,CLE00,rev,metag}.

A basic  difference between  $K^-$ production  in heavy
ion and in $N$--$N$ collisions can be attributed to 
strangeness-exchange  reactions between secondaries~\cite{ko84}
which are absent in $N$--$N$ collisions.
In heavy-ion collisions  more channels are available for
anti--kaon production, namely strangeness-exchange
processes like $\pi + Y \, \rightarrow \,  K^- + N$.
The hyperons $Y$ (i.e.  $\Lambda$ and $\Sigma$) are
produced together  with a $K^+$ since this is  energetically the
most favorable way to produce strange hadrons at low energies. These
processes are usually included in transport 
models \cite{cassing,cassing3,HOA} which show the dominance of the 
strangeness-exchange
channel. Yet, different conclusions have been drawn with respect to
the influence of the $K^-N$ potential. Some authors claim
that  one needs to reduce
the $K^-$ mass in the dense medium due to the 
attractive $K^-N$ potential~\cite{cassing},
while  others \cite{HOA}  argue that the
influence of the $K^-N$ potential is negligible as the 
antikaons  are produced at the end of the interaction chain where
the density is low.

In this Letter we compare experimental results on $K^-$
production in the energy range from 1.5 $A\cdot$GeV up to RHIC energies
with a specific assumption about the production mechanism of
$K^-$. We discuss in detail the question whether the 
experimental results support the idea of chemical equilibration of
$K^-$ in the strangeness-exchange channel. For this purpose we
study the $K^-/K^+$ ratio as function of centrality and incident
energy.

\begin{center}
\begin{figure}[h]
\epsfig{file=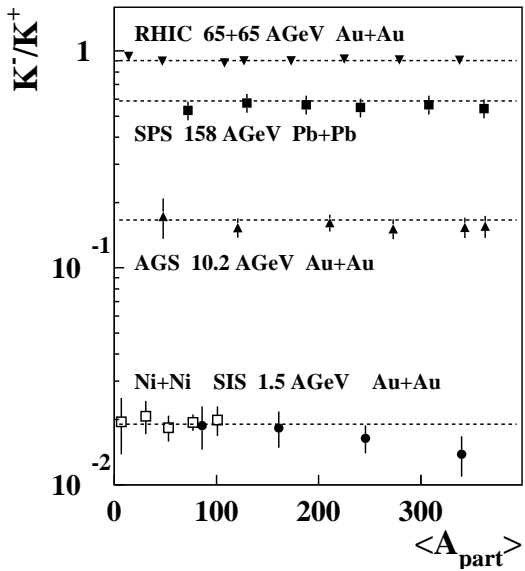,width=8cm}
 \caption{The $K^-/K^+$
ratio as a function of the number of participating nucleons $A_{part}$
obtained in heavy-ion collisions at SIS, AGS, SPS and RHIC
incident energies  \protect\cite{Ahle_2000,AF,NA49,STAR}.  The
dotted lines represent the predictions of a statistical
model\protect\cite{CLE99,CLE00,rev}.}
\label{KPKM_SIS_RHIC}
\end{figure}
\end{center}

The experimental data on kaon production in heavy-ion collisions
show that  in a very broad energy range from SIS up to RHIC  the
 $K^-/K^+$ ratio is independent of the collision centrality. This is illustrated
 in Fig.~\ref{KPKM_SIS_RHIC} for SIS, AGS, SPS and RHIC energies.
 Although at high
 incident energies this result could be expected
 since  $K^-$ is dominantly  produced together with $K^+$,
 it is rather surprising to be also valid at SIS.
Here, $K^+$ production is close to threshold (the threshold in
$N$--$N$ collisions is 1.58 GeV 
for  $K^+$ being produced together with  $\Lambda$), while $K^-$
production is far below threshold as $K^+K^-$ pair production
requires 2.5 GeV in the laboratory frame. In central heavy ion
collisions much higher densities are reached than in peripheral
processes. As the number of multiple collisions 
increases with density, one would expect the
centrality dependence of these two particles to be very different.
This expectation is in clear contrast to the data as shown in Fig.~\ref{KPKM_SIS_RHIC}.

\begin{center}
\begin{figure}
\vspace*{-2.8cm}
\includegraphics[width=6cm]{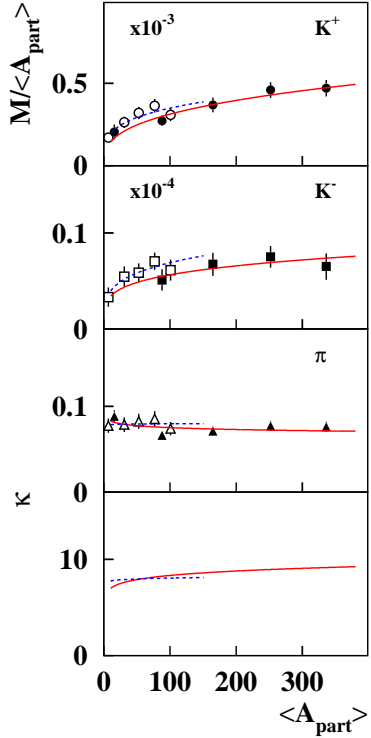}
\caption{Particle yields $M$ per participant $A_{part}$ as a
function of $A_{part}$ obtained
  from Ni+Ni (open symbols) and
Au+Au collisions  (full points) at 1.5 $A\cdot$GeV. The upper two
panels show the
multiplicity of $K^+$ and $K^-$ per $A_{part}$, 
the third panel shows the pion multiplicity $M^*(\pi)/A_{part}$.
The lines are functions  $M \propto A_{part}^\alpha$ fitted to the
data (solid lines for Au+Au, dashed lines for Ni+Ni collisions
with $\alpha \approx 1.3$). The lower panel shows the resulting
values of $\kappa$ from Eq.~(\ref{eq5}) as determined from the
fits shown above. The data are taken from~\protect\cite{AF}.}
\label{MASS_AU}
\end{figure}
\end{center}

In thermal-statistical models, the fact that the
$K^-/K^+$ ratio is independent of centrality is natural. In
 the grand-canonical description of
strangeness conservation, valid in high-energy heavy-ion
collisions,    particle densities are   independent of the
volume, thus also of the centrality  or the number of participating
nucleons $A_{part}$. In a canonical description, as 
required at SIS energies, 
strange particle densities are strongly changing with the
number of participants \cite{CLE99,CLE00,rev}. However,
this dependence enters both in the $K^+$ and in the $K^-$
production and  cancels in the ratio. A non--linear
increase of $K^+$ and $K^-$ yields as a function of the centrality
has indeed been found experimentally~\cite{AF} and can also 
be seen in the upper two panels of Fig.~\ref{MASS_AU}.

The observation of a $K^-/K^+$ ratio being independent of the
centrality (see Fig.~\ref{KPKM_SIS_RHIC}), could already be considered
as an indication   of chemical equilibration of the strangeness
production in  low-energy heavy-ion collisions. However, similar
results are also obtained in dynamical transport models as a
consequence of the interplay between in--medium effects and the nuclear
absorption of kaons \cite{cassing,cassing3}. Thus, to establish
chemical equilibration of kaons  requires further discussion which we
present below.

At low incident energies the strangeness-exchange reaction
\begin{equation}
\pi + Y \, \rightleftharpoons  \,  K^- + N. \label{eq1}
\end{equation}
plays a key role in $K^-$  production. This process  has a
low threshold, approximately 180 MeV, and a large cross
section~\cite{ko84}.

If the rates for $K^-$ production are equal to those for   $K^-$
absorption, the  reaction  in Eq.~(\ref{eq1}) is in chemical
equilibrium. In this case   the law-of-mass action is applicable
and leads to the following  relation between particle yields
\begin{equation}
\frac{[\pi] \cdot [Y]}{[K^-] \cdot [N]} \, = \, \kappa \label{eq2}
\end{equation}
with $[x]$ being the multiplicity of particle $x$ and $\kappa$ the
equilibration constant~\cite{HO_SQM2000}.

The above relation between kaon, hyperon and pion multiplicities
required by detailed balance  is a straightforward direct test of
chemical equilibration in the final state. Before the
above relation can be compared to measured yields, one needs
to take into account that  the pion density [$\pi$] in
Eq.~(\ref{eq2}) contains unequal contributions from [$\pi^+$],
[$\pi^0$] and [$\pi^-$] due to the isospin  asymmetry of the
colliding system. All available strangeness-exchange reactions
(Eq.~\ref{eq1}) can be subdivided into two groups: two channels
involving $\Lambda$ scattering of pions 
and five channels involving  $\Sigma$'s. Listed below
are the production channels with   their corresponding isospin
weight
\begin{center}
\begin{tabular}{lclr}
$\pi^0+\Lambda^0$ &$\rightleftharpoons$& $K^- + p , \, \bar{K}^0 + n $ &0.5\\
$\pi^-+\Lambda^0$ &$\rightleftharpoons$& $K^- + n                    $ &1.0\\
$\pi^-+\Sigma^+ $ &$\rightleftharpoons$& $K^- + p , \, \bar{K}^0 + p $ & 0.5\\
$\pi^-+\Sigma^0 $ &$\rightleftharpoons$& $K^- + n                    $ & 1.0\\
$\pi^0+\Sigma^0 $ &$\rightleftharpoons$& $K^- + p , \, \bar{K}^0 + n $ & 0.5\\
$\pi^++\Sigma^- $ &$\rightleftharpoons$& $K^- + p , \, \bar{K}^0 + n $ & 0.5\\
$\pi^0+\Sigma^- $ &$\rightleftharpoons$& $K^- + n                    $ & 1.0 \\
\end{tabular}
\end{center}
The relevant $\pi$ multiplicities $M^*(\pi)$ are obtained as
follows: the contribution via the $\Lambda$ channel is
$M^{\Lambda}(\pi) = 1/3 [\pi^0] + 2/3 [\pi^-]$. Correspondingly,
the channels via $\Sigma$ yield $M^{\Sigma}(\pi) = 1/7 [\pi^+] +
3/7 [\pi^0] + 3/7 [\pi^-]. $ The multiplicity of   $[\pi^0]$ is
taken as $0.5 \cdot ([\pi^+] + [\pi^-])$. The hyperons are
produced together with $K^+$ and $K^0$ in equal rates, hence, $
[Y] = [K^+] + [K^0] \approx 2 \cdot [K^+]$. The separation of the
hyperons into $\Lambda$ and $\Sigma$ baryons is  not possible experimentally.
 However, the relative yields of these particles  can
be estimated roughly using thermal occupancies. Neglecting the
isospin asymmetry and within  the limit of $m_Y/T>>1$  which is
valid at low incident energies, one gets

\begin{equation}
\frac{N_{\Sigma}}{N_{\Lambda}} \, \simeq   \,
\frac{g_{\Sigma}}{g_{\Lambda}}
\left(\frac{m_{\Sigma}}{m_{\Lambda}}\right)^{3/2}
\,\exp\left(-\frac{m_{\Sigma}-m_{\Lambda}}{T}\right). \label{eq3}
\end{equation}

The mass difference $m_{\Sigma}-m_{\Lambda} \approx 75$ MeV is
approximately equal to the freeze-out  temperature at beam
energies around (1 -- 2) AGeV \cite{CLE99,metag}. This together
with the ratio of the spin--isospin degeneracy  factors
$g_{\Sigma}/g_{\Lambda} = 3$ implies that
 $N_{\Lambda} \approx N_{\Sigma}$ is a reasonable approximation.
 Hence, the probability to form a $K^-$ with a $\Lambda$ or with a
 $\Sigma$ is the same.
The multiplicity of pions in  Eq.~(\ref{eq2}), due to the various
isospin combinations, is then obtained as
\begin{equation}
M^*(\pi) = \frac{31}{42} [\pi^-] + \frac{11}{42} [\pi^+].
\label{eq4}
\end{equation}
The multiplicity of nucleons $[N]$  in Eq.~(\ref{eq2}) can be
related to the average number of participants $A_{part}$.
Thus the law of mass action can be formulated directly
through experimentally accessible observables leading 
to 
\begin{equation}
\frac{2 \cdot [K^+]}{[K^-]} \cdot \frac{M^*(\pi)}{ A_{part}} \ =
\, \kappa .\label{eq5}
\end{equation}
This relation also holds for reactions involving  baryon exchange as  in $\Delta N\rightleftharpoons KNN$ as $\Delta$ correponds to $\pi N$.

In the two upper panels of Fig.~\ref{MASS_AU} the dependencies of
the $K^+$ and the $K^-$ multiplicities on $A_{part}$ are shown for
Ni+Ni and  Au+Au collisions at 1.5 AGeV ~\cite{AF}. The lines are
functions $M \propto A_{part}^\alpha$ fitted to the data (solid
lines for Au+Au, dashed lines for Ni+Ni) yielding similar values
for $\alpha$ between 1.3 and 1.4 for both systems and both
particle types. Consequently, the $K^-/K^+$ ratio is centrality
independent, as seen in Fig.~\ref{KPKM_SIS_RHIC}. The third panel
of figure~\ref{MASS_AU} shows that also the pion
multiplicity per $A_{part}$  is approximately constant with
centrality.

  The ratio $\kappa$ from Eq.~(\ref{eq5}) is shown  in the lowest part
of Fig.~\ref{KPKM_SIS_RHIC}. It has been calculated 
separately for Au+Au and Ni+Ni from the fitted functions 
in the panels above. 
The experimental data obtained in
Ni+Ni and Au+Au collisions at 1.5 AGeV confirm the  validity of
the law of mass action as $\kappa$ is
independent of centrality within the experimental 
uncertainties. In addition, the value of the $K^-/K^+$
ratio is in very good agreement with the statistical
model~\cite{rev,CLE00} as  shown in Fig.~\ref{KPKM_SIS_RHIC}.
These results provide evidence that at SIS energies the $K^-$
yields follow the conditions for chemical equilibrium in the final state.

\begin{center}
\begin{figure}
\vspace*{-2.8cm}
\includegraphics[width=7.5cm]{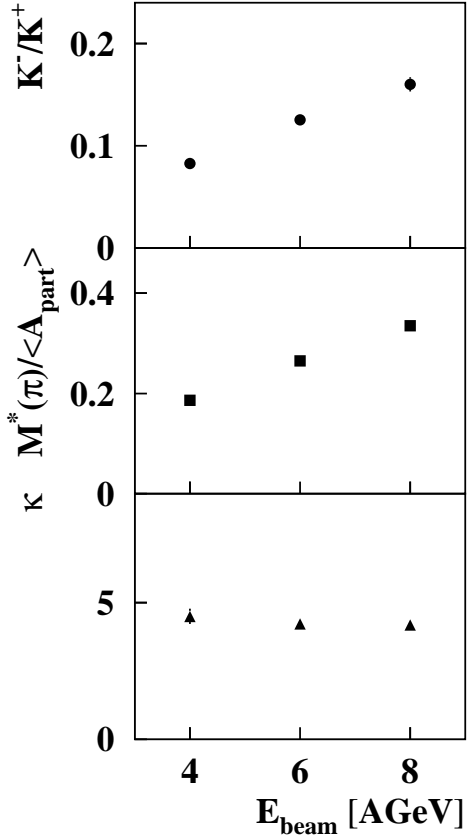}
\caption{{ The $K^-/K^+$ ratio (upper part), pion multiplicity per
participant $M^*/A_{part}$ (middle part) and the factor $\kappa$
from Eq.~(\ref{eq5})  (lower part) as a function of incident
energy  for central AuAu collisions.
The data are taken from~\cite{Ahle_2000,Klay}.}}
\label{MASS_AGS}
\end{figure}
\end{center}

Recently, new  experimental results for particle yields in central
Au+Au collisions at AGS energies have become available
\cite{Ahle_1998,Ahle_2000,Klay} and allow for further tests of chemical
equilibration of $K^-$ via the law of mass action. 
Figure~\ref{MASS_AGS} shows the $K^-/K^+$ ratios and the
pion multiplicities both increasing with incident energy.
Nevertheless, the equilibration constant $\kappa$, shown in the
lower part of Fig.~\ref{MASS_AGS}, appears to depend rather weakly
on the incident energy indicating that the $K^-$ are produced via
strangeness exchange in chemical equilibrium. Consequently, in the
energy range between SIS and AGS one expects from Eq.~(\ref{eq5})
that there should be a linear correlation between the $K^-/K^+$
and $M^*(\pi )/A_{part}$ ratios.

\begin{center}
\begin{figure} 
\vspace*{-.5cm}
\includegraphics[width=8.5cm]{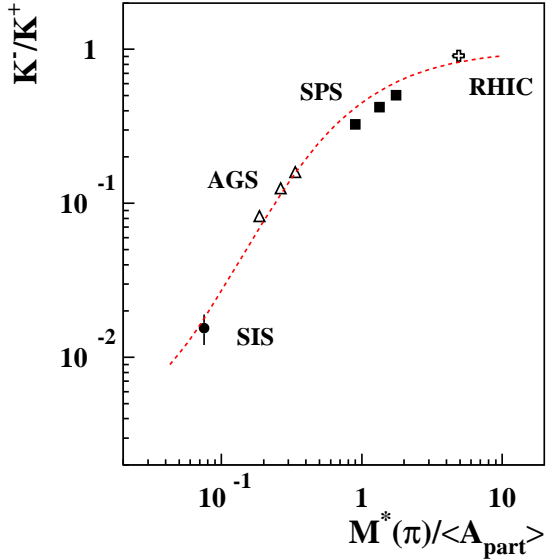}
\caption{{ The relation between the $K^-/K^+$ ratio and the pion
multiplicity $M^*(\pi)$ per participant $A_{part}$. Data are for
central Au+Au/Pb+Pb collisions at SIS, AGS, SPS and RHIC
energies~\cite{AF,Ahle_2000,Klay,STAR,NA49_energy}.  The line is
the statistical model result obtained along a unified freezeout
curve of fixed energy/particle $\simeq 1$ GeV \cite{1gev,max}. }}
\label{KPKM_PI_THERM}
\end{figure}
\end{center}

 As the results  presented above strongly point to
chemical equilibration in the strangeness exchange channel and the
$K^-/K^+$ ratio is constant as a function of the centrality, one
might be tempted to interpret the constancy also at high incident
energies by equilibration in the strangeness-exchange channel.
This is, however, not the case. Figure~\ref{KPKM_PI_THERM} shows
the relation between the $K^-/K^+$ ratio and the pion
multiplicity. As expected from the discussion above, up to top AGS
energies a linear relation between these two quantities holds. At
SPS and RHIC the $K^-/K^+$ ratio approaches unity while the pion
multiplicity keeps increasing. At these high incident energies,
the kaon production is dominated by $K^-K^+$ pair production.

Figure ~\ref{KPKM_PI_THERM} shows the relation between the
$K^-/K^+$ ratio and the pion yield per participant  in AuAu/PbPb
collisions obtained  in  the energy range from SIS up to RHIC. As
seen in Fig.~\ref{KPKM_PI_THERM} a linear relation between these
two quantities indeed holds up to AGS energies. This is a direct
evidence for chemical equilibration of $K^-$ production via the
strangeness exchange reaction. For higher collision energies at
SPS and RHIC the strangeness exchange processes are not any more
the leading reactions for the $K^-$ production. At  these incident
energies, the kaon yield is dominated by direct $K^-K^+$ pair
production and the linear relation between the $K^-/K^+$ and the
pion/participant ratios no longer holds. At SPS and RHIC the
$K^-/K^+$ ratio approaches unity while the pion multiplicity keeps
increasing. The statistical model results obtained along a unified
freeze-out curve \cite{1gev} are seen in Fig.~\ref{KPKM_PI_THERM}
to be consistent with data in the whole energy range. The change
in slope seen in Fig.~\ref{KPKM_PI_THERM} appears for incident
energies between 20--40 $A\cdot$GeV where the composition of the
collision fireball changes from baryonic to mesonic \cite{max}.

In summary, we have shown that in low energy heavy-ion collisions
at SIS and AGS energies the data confirm the chemical equilibration of
$K^-$ production. In this energy range the dominant process
for $K^-$ production is due to the
strangeness exchange channels $\pi
+ Y \, \rightleftharpoons \,K^- + N$. These channels link the
$K^-$ production to the yield of the $K^+$ production in
$NN \rightarrow K^+ \Lambda N$ reactions. If strangeness exchange
appears in chemical equilibrium then according to the law of mass
action the $K^-/K^+$ ratio  scales with the number of pions per
participant. This scaling could be observed in two different
ways: 
\begin{itemize}
\item the $K^-/K^+$ ratio should be independent of centrality
since $<\pi>/A_{part}\simeq const.$  and 
\item with increasing
incident energy an increase of the $K^-/K^+$ ratio should be
proportional to  $<\pi>/A_{part}$. 
\end{itemize}
Both of these chemical
equilibrium conditions are well consistent with the data up to AGS
energies. At SPS and RHIC energies the condition (ii) is violated
since the dominance of the strangeness-exchange reactions is
replaced by direct $K^+K^-$ pair production which, however still
preserves the centrality independence of the $K^-/K^+$ ratio.


 We acknowledge stimulating discussions with P. Braun-Munzinger and J. Stachel
and the support of the German Bundesministerium f\"ur Bildung und
Forschung (BMBF), the Polish State Committee for Scientific
Research (KBN) grant 2P03 (06925) and the 
National Research Foundation (NRF, Pretoria) and the URC of the
University of Cape Town.

\end{document}